# Efficient terahertz emission from InAs nanowires

Denis V. Seletskiy,[1,4,*] Michael P. Hasselbeck,[1] Jeffrey G. Cederberg,[2] Aaron Katzenmeyer,[3] Maria E. Toimil-Molares,[3] François Léonard,[3] A. Alec Talin[3,†] and Mansoor Sheik-Bahae[1]

[1]*Department of Physics and Astronomy, University of New Mexico, Albuquerque, NM 87131*
[2]*Sandia National Laboratories, Albuquerque, NM 87185*
[3]*Sandia National Laboratories, Livermore, CA 94551*
[4]*Air Force Research Laboratory, Space Vehicles Directorate, Kirtland AFB, NM 87117*
[†]*Center for Nanoscience and Technology, National Institute of Standards and Technology, Gaithersburg, MD 20899*
[*]e-mail: d.seletskiy@gmail.com

**Abstract**
We observe intense pulses of far-infrared electromagnetic radiation emitted from arrays of InAs nanowires. The THz radiation power efficiency of these structures is about 15 times higher compared to a planar InAs substrate. This is explained by the preferential orientation of coherent plasma motion to the wire surface, which overcomes radiation trapping by total-internal reflection. We present evidence that this radiation originates from a low-energy acoustic surface plasmon mode of the nanowire. This is supported by independent measurements of electronic transport on individual nanowires, ultrafast THz spectroscopy and theoretical analysis. Our combined experiments and analysis further indicate that these plasmon modes are specific to high aspect ratio geometries.

Manipulation of electromagnetic fields at sub-wavelength dimensions is an emerging trend in nanoscience and nanotechnology. This is realized by modification and control of material properties and geometries on the nanoscale [1]. For instance, the response of free carriers to external perturbations gives rise to modified behavior of slab plasmon and plasmon-polariton modes in confined geometries [2]. The conversion of electromagnetic waves into propagating charge carrier waves allows for geometric control, localization [3] and guiding of light [4] at sub-wavelength dimensions; this has provided the impetus for the emerging field of plasmonics or metal optics. A particularly interesting area is that of plasmonics in the Terahertz (THz) regime, where for example, recent research on the excitation of surface plasmon modes in tapered metal wire geometries [5-7] demonstrated imaging of electronic characteristics of materials on micro- and nano- scales [7,8].

Improvement of the time-resolved nanoscale imaging at THz frequencies requires an efficient local emitter/detector combined with broadband time-domain spectroscopy methods [9,10]. Nanostructured semiconductors in combination with recent techniques for high power ultrafast THz excitation sources [11-14] can serve as intense THz nanoscale emitters. Nanowires have been the subject of extensive ultrafast spectroscopic studies at the visible and near infrared frequencies [15]. Recent experiments characterized the far-infrared optical response of GaAs nanowires using ultrafast THz spectroscopy [16]. The emission of THz pulses has been reported from a variety of nanostructures [17-23], but the nature of the ultrafast excitation mechanism is still not well understood [17-20]. In this work, we show that InAs nanowires can serve as highly efficient radiators of THz frequency electromagnetic transients. We argue that nanowires overcome the problem of internal reflection that occurs at the air-semiconductor macro-scale interface, leading to much higher emission efficiency compared to planar substrates. Analysis of the dielectric function including transverse confinement suggests that the emission originates from a low-energy surface plasmon-polariton.

InAs nanowires are grown by metal-organic vapor phase epitaxy (MOVPE) using the vapor-liquid-solid (VLS) technique. The VLS growth mechanism for InAs nanowires has been thoroughly investigated [24,25]. Our version of this process uses a Si-doped, n-type GaAs wafer with a (111) surface orientation as the starting substrate. A 1 nm thick gold deposition is made on the (111)B surface by electron beam evaporation. Upon annealing under $AsH_3$ at 600°C, the gold deposit coalesces to form nanoparticles that represent a collector for InAs growth precursors. For nanowire growth, the temperature is reduced to 400°C and trimethyl indium (TMIn) is introduced into the chamber. The nanowires orient vertically as shown in the scanning electron microscopy (SEM) image in Figure 1a. Their length is in the range 10--20 µm, the diameter tapers from 450 nm at the base to 60 nm at the tip, and the surface coverage is ≈ 0.2 $µm^{-2}$. A gold nanoparticle remains attached to the tip of the nanowire after growth. In addition to nanowire growth, a planar wetting layer forms due to competitive vapor-solid heteroepitaxy on the GaAs(111)B surface, as shown in Figure 1b. Wetting layer growth occurs at a rate smaller than the vertical nanowire growth rate, leading to a thin film. We estimate a maximum wetting layer thickness of 200 nm if all the TMIn is consumed in this pathway. The presence of nanowires implies a wetting layer thickness < 200 nm.

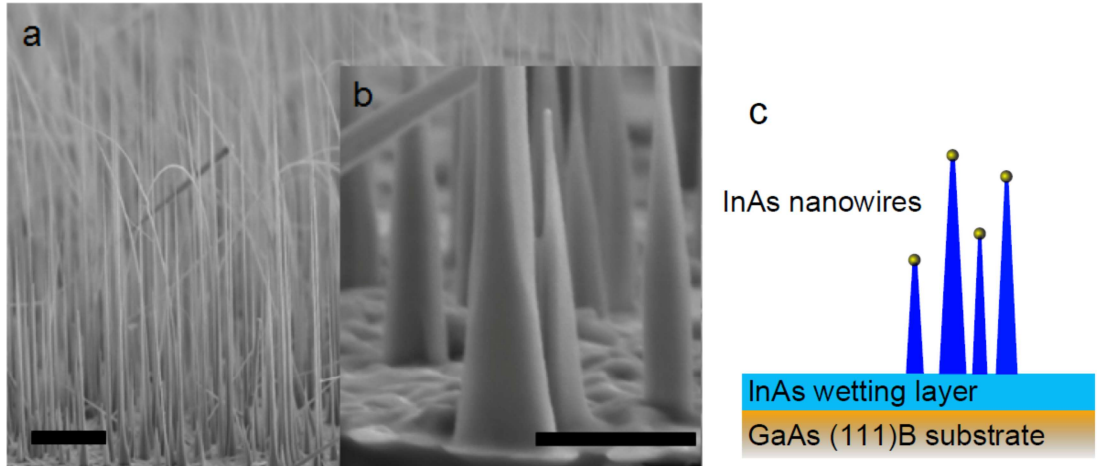

**Figure 1:** a) Low magnification cross-sectional SEM image of the InAs nanowire ensemble highlighting their high aspect ratio and their taper. The scale bar is 3 µm across. b) Higher magnification SEM image shows the base dimension and existence of a planar wetting layer formed by vapor-solid epitaxy. The scale bar is 1 µm across. c) Sketch (not to scale) illustrates the three regions of the sample that are potential sources of THz radiation transients. Our experiments show that the signal from the InAs nanowires dominates the signal from the InAs wetting layer or the Si-doped GaAs substrate.

We excite the InAs nanowires using ultrashort pulses from a mode-locked Ti:Sapphire laser oscillator (average pulse energy 5 nJ, duration 60 fs) with wavelength centered at 820 nm. The p-polarized pulses are focused on the samples (~ 400 $nJ/cm^2$) at an incidence angle of 45° with respect to the substrate surface. We characterize the emitted electromagnetic transients using a linear autocorrelation, where two identical pulses are delivered to the sample in a Michelson interferometer geometry [26-28]. The THz radiation is collected in the specular direction and imaged onto a liquid-He cooled Si-bolometer with a pair of off-axis parabolic mirrors. A Teflon filter is inserted into the THz beam path (unpurged laboratory air) to remove background laser light. Linear autocorrelations and the corresponding power spectra (Fourier transforms) are presented in Fig. 2. The amplitude of the signal (at zero delay) obtained with the nanowires [Fig.(2a), top] is only about 2 times weaker than a slightly n-doped InAs substrate [Fig.2a, middle], despite the fact that the fill factor of the wires is only of the order of 0.03. This shows that nanowires are unusually good emitters of THz radiation, with an estimated radiation enhancement of ~ 15 in emitted power, compared to the InAs substrate.

To verify that the emitted radiation originates from the nanowires, we test a control sample with the nanowires removed: a thin (less than 200 nm) InAs wetting layer on the same GaAs substrate used for nanowire growth. To remove the nanowires, we place the sample in an ultrasonic bath and have verify their absence with an optical microscope. This control sample is essential to isolate the contribution of the InAs wires from the InAs wetting layer, which builds up during the growth process. The control sample shows a very weak signal [Fig. (2a), bottom], indicating InAs nanowires are the source of strong THz radiation. (The emission peak at ~ 3.5 THz demonstrates the broadband response of the detection system.)

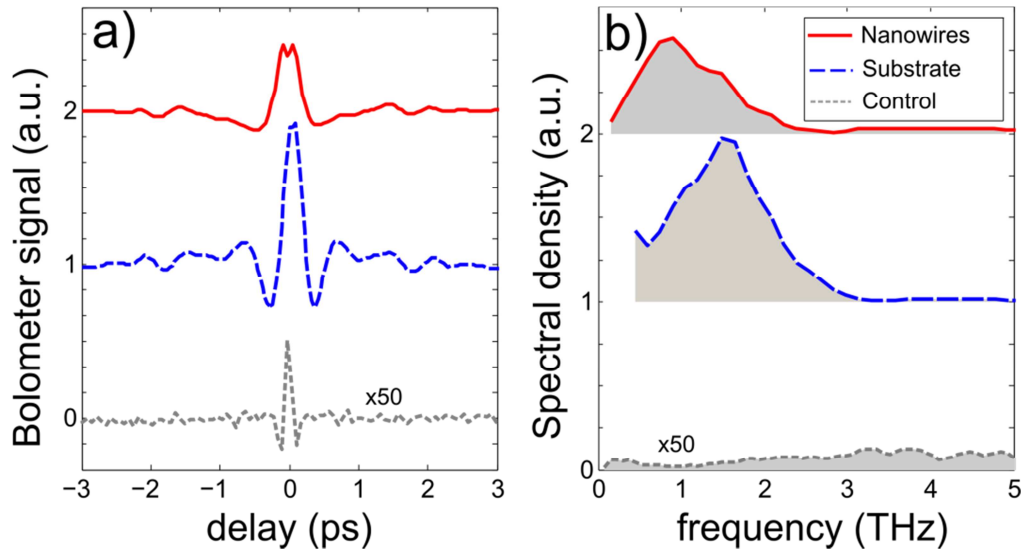

**Figure 2:** Autocorrelation signals (a) and corresponding Fourier spectra (b) for an InAs nanowire sample (top), an InAs substrate (middle), and control sample (bottom) as described in the text. The top two signals are plotted on the same relative scale; the bottom signals are multiplied 50x. Curves are offset vertically for clarity.

To understand the large enhancement of the THz emission in nanowires compared to the bulk material, we must first discuss the origin of the THz radiation. This has two components: the initial excitation mechanism that leads to carrier acceleration, and the subsequent radiation. In bulk semiconductors, ultrashort optical excitation can lead to dynamical charge transport of photo-carriers that drives emission of electromagnetic radiation [26,29]. Excitation of the semiconductor above the direct energy band gap generates electron-hole pairs at a penetration depth determined by the inverse absorption coefficient. In the case of InAs nanowires, we consider four different physical mechanisms: i) acceleration in a surface depletion field, ii) acceleration due to fields at the gold-nanoparticle/nanowire interface, iii) ambipolar diffusion, i.e. the photo-Dember effect [29,30] and (iv) nonlinear mechanisms such as optical rectification (OR) or shift currents [31]. The first mechanism is known to be important in n-doped GaAs and InP where the surface depletion field is large (> 10 kV/cm) [32,33]. Narrow gap semiconductors such as Te, InSb, and InAs have negligible surface fields, making this mechanism unimportant here [30]. The second mechanism is also insignificant because the Fermi level is pinned at the conduction band minimum in InAs, resulting in a negligible Schottky barrier and band bending [34]. If a Schottky barrier did exist, the associated depletion region would only account for a tiny volume fraction of each nanowire.

To differentiate between the photo-Dember and the nonlinear processes, we note that the power of THz pulses originating from OR or shift currents would increase quadratically with laser excitation at small conversion efficiencies, and saturate at high intensities. The power dependence of the photo-Dember effect can be complicated because transport depends critically on carrier density through scattering and screening. For bulk systems and under various excitation conditions, the dependence of the radiated THz power on the excitation levels has been observed to be either linear [29,35] or quadratic [36,37] at low power, while exhibiting saturation behavior at higher powers [35-37]. Figure 3 shows that the emitted power increases linearly with excitation intensity as the pumping level varies by over 2 orders of magnitude. This rules out any meaningful contribution from OR or shift currents, leaving only the photo-Dember mechanism as a plausible source for the initial carrier acceleration. We also point out that the observed emission was largely independent of the polarization state of the excitation, in accord with the photo-Dember mechanism and in contrast to that expected from the nonlinear contributions.

While the photo-Dember effect can directly lead to radiation, it is also well-known that charge carriers will move to screen the photo-carrier distribution. If the perturbation occurs on a time shorter than the inverse cold plasma frequency and the rate of momentum relaxing collisions is sufficiently small, coherent collective motion of cold charge carriers (plasmons) will take place leading to a macroscopic dipole radiating close to the plasma frequency [38,26]. Experiments with bulk InAs and GaAs planar layers identified cold plasmon, phonon, and mixed-mode components in the THz radiation spectra [26,28,39]. These longitudinal modes are poles of the polarization and must be included in the analysis of the free-space radiation spectra, regardless of the exact nature of the excitation mechanism [40].

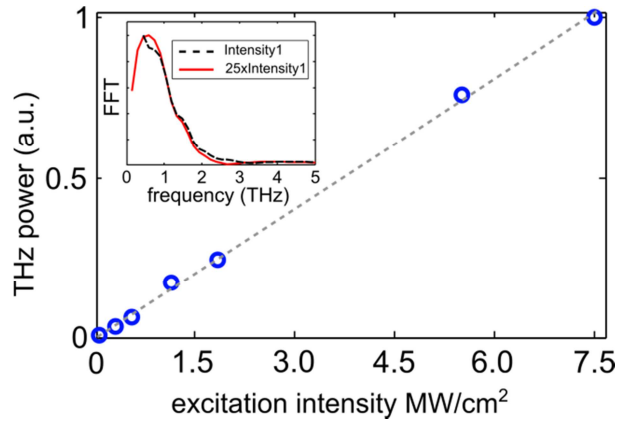

**Figure 3:** THz power changes linearly as excitation intensity is varied (via variable neutral density filter) over a range of 150. Inset shows the normalized spectra of THz transients from InAs nanowires as the intensity varies by a factor of 25. Absence of a spectral shift is consistent with radiation emanating from coherent plasma.

We assume that the radiation originates from plasmons, and use a classical Drude response [38] to analyze the spectra in Fig. 2. The simplest approach is to first consider the case of *bulk* plasmons. The plasmon is excited at frequency $\omega = \sqrt{\omega_p^2 - \gamma^2/4}$, where $\gamma$ is the damping rate (i.e. momentum relaxation time) attributed to electrons scattering with longitudinal-optical (LO) phonons and $\omega_p$ is the Langmuir plasma frequency:

$$\omega_p = \sqrt{\frac{n_e e^2}{m^* \varepsilon_0 \varepsilon_\infty}} \quad (1)$$

Here, $n_e$ is the electron density, $\varepsilon_0$ is the permittivity of free space, $m^* = 0.023 m_e$ is the electron effective mass in the conduction band and $\varepsilon_\infty = 11.8$ for InAs. Assuming an electron-donor ion plasma (i.e. neglecting holes), we estimate an electron density for the wire ensemble to be ~ $5 \times 10^{15}$ cm$^{-3}$ based on the frequency of the peak maximum and its broadening [Fig. (2b)].

To test whether the bulk plasmon theory explains our results, we independently measure the carrier concentration by performing electronic transport measurements. This is accomplished with two independent methods: (1) probing individual nanowires directly on the growth substrate inside of a SEM [41,42] and (2) fabricating individual nanowire field effect transistor (FET) devices. Individual nanowires are contacted with a tungsten nanoprobe retrofitted inside of a SEM. A large area backside Ohmic contact completes the electrical circuit [Fig. 4(a)]. The high aspect ratio geometry of the nanowires leads to reduced carrier screening and the onset of space-charge limited current (SCLC) at lower voltage than would be expected for a bulk, low-aspect ratio specimen with similar free carrier concentration [43]. The symmetrical nature of the I-V curve in Fig. 4(b) afforded by the two Ohmic [41,42] contacts despite their geometrically asymmetrical nature, is consistent with SCLC (a bulk limited regime). This is in marked contrast to rectifying I-V characteristics observed for Au catalyst/Ge nanowire contacts where a high Schottky barrier is formed [44]. Transport is diffusive in SCLC so for a specimen with a nanowire geometry, the mobility can be calculated directly using the relation [43]

$$\mu = \frac{L I_{SCLC}}{\pi \varepsilon V^2} \quad (2)$$

where $L$ is the nanowire length, $I_{SCLC}$ is the current and $V$ is the applied bias. Once the mobility is known, the carrier concentration can be extracted from the conductivity (measured at low bias).

Alternatively, $n$ can be determined using the crossover voltage $V_c$ between the ohmic and SCLC regimes. Accounting for SCLC geometrical scaling in nanowires [43,45] we have

$$n_e = \varepsilon V_c / e R^2, \quad (3)$$

where $R$ is the nanowire radius.

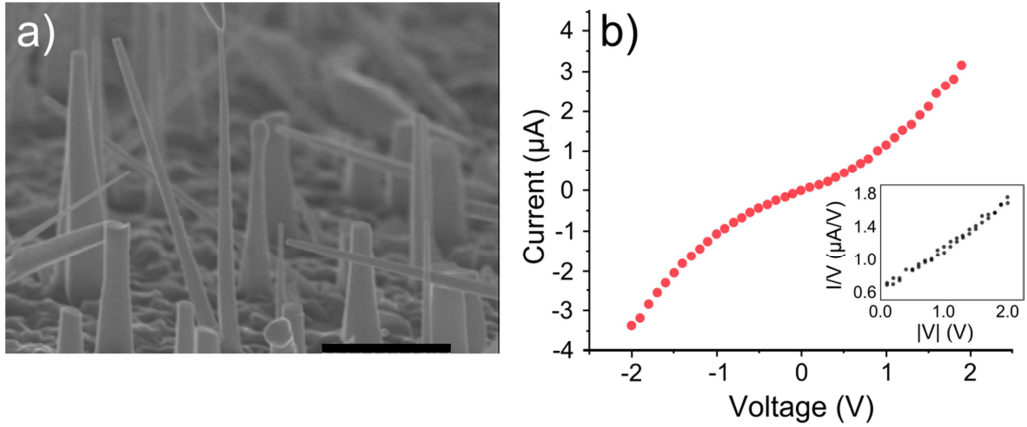

**Figure 4:** (a) SEM image depicting the probing of an individual nanowire on the growth substrate, scale bar corresponds to 1.5 $\mu$m; (b) I-V curve for the InAs nanowire measured in (a). (Inset) The same data plotted in the formalism of space-charge-limited current enables one to extract the Ohmic (intercept) and space-charge-limited (slope) contributions to the total current and hence the cross-over voltage.

In the second experiment, source and drain Ti/Au electrodes are fabricated over randomly dispersed nanowires on a n-Si/SiO$_2$ substrate. An example of one of the nanowires along with its electrical transport data is shown in Fig. (5). We calculate the coupling capacitance $C$ between the InAs nanowire and the back-gate through the 100 nm thick SiO$_2$ gate dielectric using the cylinder-on-plate model. Mobility and carrier concentrations are then extracted from the transconductance using the relations $\mu = \dfrac{L^2}{V_{SD}C}\left(\dfrac{dI}{dV_G}\right)$ and $n_e = CV_{th}/e$ where $V_{SD}$, $V_G$, and $V_{th}$ are the source-drain, gate and threshold voltage, respectively. Both measurement techniques show that as the nanowire diameter increases from ~40 nm to ~250 nm, the carrier concentration decreases from ~10$^{18}$/cm$^3$ to ~10$^{16}$/cm$^3$ while the mobility increases from ~100 cm$^2$/Vs to 3000 cm$^2$/Vs [41]. These trends are in reasonable agreement with other measurements on VLS grown InAs nanowires using a FET geometry [46]. It must be pointed out, however, that nanowire mobility measurements provide lower-limit estimates of the intrinsic mobility, since Ohmic losses at the contacts as well as surface defects introduced during processing of the NWs tend to lower the measured values.

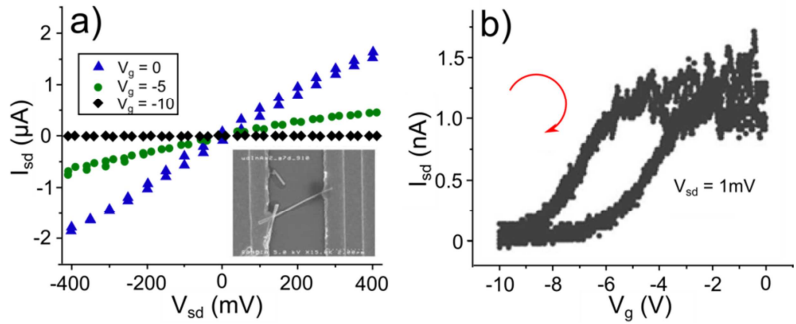

**Figure 5:** (a) Source-drain current vs. source-drain voltage sweeps of a representative FET at different fixed gate biases showing the n-type depletion mode behavior. (b) Transfer characteristic of the FET shown in (a) which enables determination of the mobility and carrier concentration by way of the transconductance and threshold voltage, respectively. The extracted device parameters are $n_e = 4.8 \times 10^{17}$ cm$^{-3}$ and $\mu = 101$ cm$^2$/V·s. Nanowire dimensions are R = 34 nm, L = 3 μm.

Our electronic measurement of the carrier concentration differs by nearly two orders of magnitude from that determined by the bulk model [Eq. (1)], suggesting that the bulk model is not applicable to the nanowire geometry. To resolve this discrepancy, we consider a more realistic description of the longitudinal polarization modes for the cylindrical nanowire geometry by applying a formalism developed by Pitarke et. al. [47]. This theory provides the dielectric function of a conducting cylinder yielding eigenmodes of the geometrical structure as a function of parameter $q_\ell R$ and an integer mode number $m$, arising from the cylindrical symmetry. The dimensionless parameter $q_\ell R$ is the product of the wavevector component along the nanowire axis ($q_\ell$) and the nanowire radius $R$. We modify the model to consider discrete axial modes with $q_\ell = \ell\pi/L$, where $L$ is the length of the nanowire and $\ell > 0$ is an integer [48,49]. Figure (6a) shows the dispersion relation of the modes for the case of the assumed local response [47]. Note that in addition to the dispersion-less bulk mode at $\omega_p$, there exist surface plasmon modes indexed by the integer $m$ and given by [47,50]:

$$\omega(m, q_\ell R) = \omega_p \sqrt{\frac{\varepsilon_\infty}{\varepsilon_\infty - \alpha_m(q_\ell R)\varepsilon_0}} \quad (4)$$

where $\alpha_m(x) = I_m'(x)K_m(x)/K_m'(x)I_m(x)$, $I_m$ and $K_m$ are modified Bessel functions and the prime indicates the derivative of the corresponding functions. In the limit of $q_\ell R \gg 1$ ($\alpha_m \to -1$), the surface plasmon modes approach the planar surface plasmon frequency $\omega_{sp} = \omega_p\sqrt{\varepsilon_\infty/(\varepsilon_\infty + \varepsilon_0)}$ (Fig 6a), which reduces to the familiar $\omega_{sp} = \omega_p/\sqrt{2}$ for $\varepsilon_0 = \varepsilon_\infty = 1$. The $m=0$ modes exhibit *acoustic-like* dispersion for small $q_\ell R$, while the higher modes ($m > 0$) are essentially dispersion-less around the $\omega_{sp}$ frequency. Thus, for the assumed purely longitudinal excitation, the only excitable radially-symmetric $m=0$ mode is considered in the remainder of our analysis.

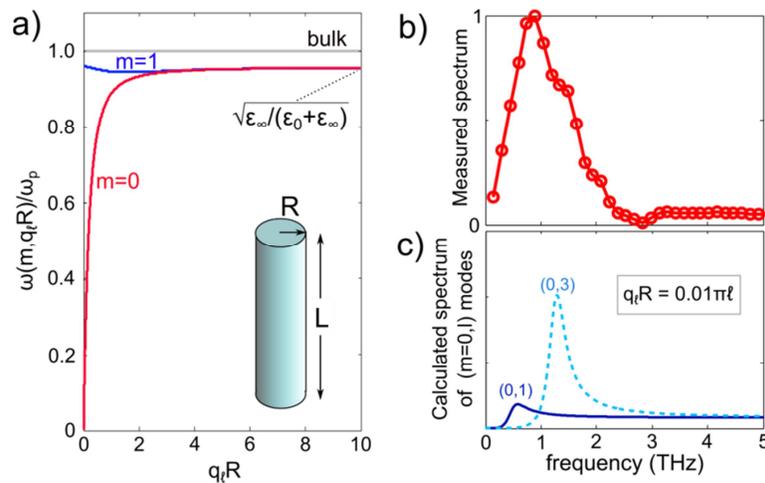

**Figure 6:** (a) Dispersion of the longitudinal bulk and surface plasmon modes; (b) Normalized measured emission spectrum of the InAs nanowire ensemble; (c) Theoretical spectra of the lowest order axial surface acoustic plasmon modes of the nanowire with $n_e = 4.8 \times 10^{17}$ cm$^{-3}$, $R = 34$ nm, $L = 3$ μm and plasmon broadening of $0.05\omega_p$.

To analyze the radiated spectra of the plasmon modes, we consider the induced polarization by an externally applied longitudinal electric field $E_{ext}$ (such as photo-Dember) given by [49]:

$$P(q,\omega) = \varepsilon_{\infty}\left(1 - \frac{1}{\varepsilon(q,\omega)}\right) E_{ext}(q,\omega), \quad (5)$$

where $\varepsilon(q,\omega)$ is the wavevector and frequency-dependent dielectric function of the wires [47]. The radiated far-field is $E_{rad} \propto \partial^2 P / \partial t^2$, so ignoring propagation effects and assuming broadband excitation, the radiated power spectrum $S_{rad}(q,\omega) \propto |1 - \varepsilon^{-1}(q,\omega)|^2 \omega^4$. Figure (7c) shows two such spectra for axially-symmetric $\ell = 1, 3$ modes taking $R/L \sim 0.01$ of the nanowire shown in Fig. (5). In accordance with antenna theory, only the axially-symmetric modes ($l$=odd) interfere constructively in the far-field [47,48]. The frequency axis in Fig. (7c) is scaled by the bulk plasmon frequency of ~12 THz as determined from Eq. (1) for $n = 4.8 \times 10^{17}$ cm$^{-3}$. The predicted acoustic-plasmon frequency of the ($m$=0, $\ell$=1) mode in this nanowire geometry is within a factor of two of the experimentally observed value [Fig. (7b)]. The next lowest axial mode shows even better overlap with the measured spectrum. This is in stark contrast to the factor of 100 discrepancy obtained from the bulk-plasmon model analysis [Eq. (1)] and suggests that the observed THz emission originates from a low-energy acoustic-plasmon mode of the nanowires. This interpretation is also consistent with independent measurements of electrical transport and ultrafast THz spectroscopy.

We point out that the experimentally observed THz spectral shape is inhomogeneously broadened due to the ensemble of nanowires of varying geometry and carrier concentration. Variations in the aspect ratio ($R/L$) would result in a proportional linear shift of the frequency of the acoustic modes [Fig. (6a)], while the carrier density variations would correspond to quadratic frequency shifts. Averaging over the nanowire ensemble smears out the spectrum as shown in Fig. (6b) and prevents observation of distinct features associated with an individual nanowire [Fig. (6c)].

Our experimental results and modeling suggest the presence of a low-energy acoustic surface plasmon mode is responsible for THz emission from InAs nanowires. Surface plasmon cavity modes in metallic nanowires have been recently inferred by the near-field microscopy [51] and optical loss spectroscopy [48]. Our data indicates that we observe radiation directly from the acoustic surface plasmon modes in the Hertzian dipole geometry. These results are corroborated by a recent observation of THz emission from Si nanowires, where the role of surface modes has also been discussed [20]. We also point out that we were not able to detect any THz emission from much shorter ($L \sim 300$ nm) nanowires [18], where an appreciable momentum mismatch cannot be compensated by the lowest axial modes even in the Hertzian dipole approximation. Efficient Landau damping of plasmons is also likely to exist in short wires, further reducing THz emission [52].

Finally, we address the observed large enhancement of the emitted radiation from nanowires in comparison to a planar substrate. For an InAs substrate, coherent plasmons are excited below the surface leading to charge oscillations along the direction of the surface normal and a concomitant emission of radiation into free space. The $sin^2(\theta)$ dipole radiation pattern is sketched in Figure

(7a), where $\theta$ is the internal angle with respect to the surface normal. The radiation pattern is primarily oriented parallel to the semiconductor surface. When propagating from a high refractive index semiconductor (n ~ 4) into free-space, Snell's Law at the interface must be satisfied or total internal reflection will occur. Out-coupling from the dipole takes place in a narrow escape cone with an apex angle of $\theta_c = \arcsin(1/n) \approx 15°$ sketched as the shaded triangle in Fig. (7a). Light rays striking the surface at angles θ greater than the apex angle of the escape cone suffer total internal reflection and remain trapped in the semiconductor. Only a small portion of the dipole radiation pattern (< 1%) has the proper wave-vector to satisfy coupling to free-space [30]. We note that the refractive index is a complex quantity and can be highly dispersive in the vicinity of the plasma and optical phonons. Plasma oscillations are strongly damped in InAs, however, so the complex index is relatively constant at and above 1 THz.

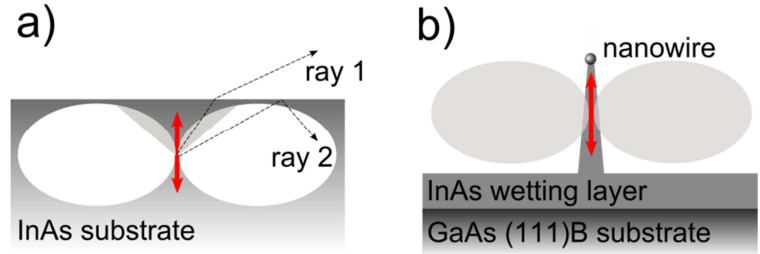

**Figure 7:** (a) Radiation pattern from a plasma dipole (oscillation direction indicated by bold double arrow) beneath the surface of a planar semiconductor. The shaded region within the lobes depicts the radiation escape cone; only light within this cone (ray 1) can escape into free space. The remaining radiation (ray 2) is trapped due to total internal reflection; (b) Schematic of coherent plasmon motion and light propagation with a free-standing semiconductor nanowire.

The radiation trapping problem is well understood and has led to the development of different schemes for improving the extraction efficiency. For example, the direction of charge carrier motion has been altered by the Lorentz force [53] where a large dc magnetic field (~ 1 T) rotates the dipole to preferentially orient one lobe of the radiation pattern in the escape cone [30,53]. Another approach is to implement structured surfaces such as low absorption, index-matched prism stripes [54] or metal edges [55]. In contrast, free-standing, vertically oriented nanowires possess an inherent geometric efficiency compared to substrates because their primary surface is parallel to the direction of collective charge motion. The effect is sketched in Fig. (7b). We estimate the enhancement of the NW emission over the substrate by calculating the ratio of emitted power into the full solid angle from NWs ($P_{NW}$) to that emitted (and coupled to free-space) from the substrate ($P_{Bulk}$):

$$\frac{P_{NW}}{P_{Bulk}} = \frac{\int d\Omega \sin^2(\theta)}{\int d\Omega \sin^2(\theta)(1-R_{\parallel}(\theta))}, \qquad (6)$$

where the assumed source is a Hertzian dipole emitter ($\propto \sin^2(\theta)$) and $d\Omega = \sin\theta d\theta d\phi$ is an infinitesimal solid angle in spherical geometry. We account for the standard Fresnel losses $(1-R_{\parallel}(\theta))$ [56] when considering emission from photo-Dember excitation (i.e. p-polarized by the nature of dipole orientation) in the substrate (denominator of Eq. (6)). Since NW volume is much smaller than the cube of the characteristic THz wavelength, its effective refractive index is near unity and hence no Fresnel losses are considered (numerator). The integration over θ in the case of a substrate is performed from 0 to $θ_c$ to account only for the escaped emission. Using Eq.(6) for index n ~ 4 the estimated enhancement factor is ~ 20 and is practically unchanged even when the experimental collection geometry (45 degree incidence with 10 degree collection half-angle) is taken into the account. The observed emission enhancement per NW of a factor of 15 compares favorably with this estimate.

An enhancement of THz emission has been observed in several other material systems. A recent experiment with opaque/blackened silicon showed that surface roughness in the form of needles (length ~2 μm, diameter ~ 300 nm) enhanced the THz output compared to an untreated surface [19]. This was ascribed to increased absorption of pump laser photons via multi-path scattering. The emission mechanism was attributed to an acceleration of photo-carriers in the surface depletion field region [19]. Radiation power from blackened silicon, however, was ~ 50 times weaker than a polished InAs substrate. THz enhancement was also observed with InN nano-pillars for a surface number density that is ~ 250 times larger than in the present work [17]. Our experiments show that InAs nanowires generate signal levels comparable to substrates, not even accounting for the spatial fill factor. This performance makes InAs nanowires highly attractive for nanoscale THz emitter and detector applications.

In summary, we have demonstrated strong emission of THz radiation from a cold plasma in InAs nanowires. There is a 15-fold enhancement in power of outcoupling efficiency compared to a bulk InAs planar substrate, which we attribute to a more favorable geometry. A geometry-dependent dielectric function that includes a low-frequency acoustic surface plasmon mode provides a good description of the observed THz radiation spectra and order of magnitude agreement with separate electronic measurements. A systematic study of the THz emission from *single* nanowires as a function of geometry must be performed to better understand plasmon dispersion in these structures.

**Acknowledgement.** DVS acknowledges useful discussions with Dr. J.N. Heyman. Partial support is provided by the Defense Threat Reduction Agency Grant DTRA01-03-D-0009-0026 and NSF Grant 0722622-MRI. Work performed in part at the U.S. Department of Energy, Center for Integrated Nanotechnologies, at Los Alamos National Laboratory, and Sandia National Laboratories. Sandia is a multiprogram laboratory operated by Sandia Corporation, a Lockheed Martin Company, for the United States Department of Energy under contract DE-AC04-94-AL85000. Research in part was performed while DVS held a National Research Council Research Associateship Award at Air Force Research Laboratory.